# The weakening relationship between the Impact Factor and papers' citations in the digital age

George A. Lozano[†], Vincent Larivière[§] and Yves Gingras[*]

[†] dr.george.lozano@gmail.com

Observatoire des Sciences et des Technologies (OST), Centre Interuniversitaire de Recherche sur la Science et la Technologie (CIRST), Université du Québec à Montréal, CP 8888, Succ. Centre-Ville, Montréal, QC. H3C 3P8 (Canada)

[§] vincent.lariviere@umontreal.ca

École de bibliothéconomie et des sciences de l'information, Université de Montréal, C.P. 6128, Succ. Centre-Ville, Montréal, QC. H3C 3J7 (Canada)

and

Observatoire des Sciences et des Technologies (OST), Centre Interuniversitaire de Recherche sur la Science et la Technologie (CIRST), Université du Québec à Montréal, CP 8888, Succ. Centre-Ville, Montréal, QC. H3C 3P8 (Canada)

[*] gingras.yves@uqam.ca

Observatoire des Sciences et des Technologies (OST), Centre Interuniversitaire de Recherche sur la Science et la Technologie (CIRST), Université du Québec à Montréal, CP 8888, Succ. Centre-Ville, Montréal, QC. H3C 3P8 (Canada)

**Abstract**

Historically, papers have been physically bound to the journal in which they were published but in the electronic age papers are available individually, no longer tied to their respective journals. Hence, papers now can be read and cited based on their own merits, independently of the journal's physical availability, reputation, or Impact Factor. We compare the strength of the relationship between journals' Impact Factors and the actual citations received by their respective papers from 1902 to 2009. Throughout most of the 20th century, papers' citation rates were increasingly linked to their respective journals' Impact Factors. However, since 1990, the advent of the digital age, the strength of the relation between Impact Factors and paper citations has been decreasing. This decrease began sooner in physics, a field that was quicker to make the transition into the electronic domain. Furthermore, since 1990, the proportion of highly cited papers coming from highly cited journals has been decreasing, and accordingly, the proportion of highly cited papers not coming from highly cited journals has also been increasing. Should this pattern continue, it might bring an end to the use of the Impact Factor as a way to evaluate the quality of journals, papers and researchers.

## Introduction

The Impact Factor (IF) was originally devised in the 1960s to guide academic libraries in their journal purchases (Archambault & Larivière, 2009). Although several other types of citation-based measures of journal impact have been recently devised, such as the SCImago Journal Rank (González-Pereira, Guerrero-Bote, & Moya-Anegón, 2009), the Eigenfactor (West & Bergstrom, 2010) and the source normalized impact per paper (Moed, 2010), the 2-year IF compiled by Thomson Reuters is still the most widely used. The IF of a given journal for a given year is defined as the mean citation rate, during that given year, of the papers published in that journal during the previous 2 years. For example, a journal's IF for 2011 considers citations received in 2011 by papers published in that journal during the years 2009 and 2010. Thus, the citation window for individual papers ranges from 1 year to almost 3 years, with an average of 2 years.

Over the past few decades, IFs have slowly permeated into the collective consciousness of scientists, and IFs have become self-reinforcing measures of journal quality, the papers therein, and their authors. Researchers now consider IFs when choosing their publication outlets; journal editors formulate policies explicitly designed to improve their IFs, and publishers advertise their IFs on their web sites. IFs are often used as a surrogate for the actual number of citations a paper recently published might eventually receive. Such a proxy might be partially justified given that, independently of the quality of the paper, a journal's IF is positively linked with the citations received by its papers (Larivière & Gingras, 2010). Since the early 1990s, as citation data became electronically available, interest and use of the IF has increased, and scholarly articles on the IF have increased exponentially (Archambault & Larivière, 2009).

The digital age also brought forth another change. Since the creation in 1665 of the *Journal des Sçavans* and the *Philosophical Transactions of the Royal Society*, which are considered to be the first two scientific periodicals, researchers have mostly read actual printed journals, so papers published in high profile journals with high circulation had a greater chance of being read and cited than papers published in less widely available journals. Now that scientific information is disseminated electronically, researchers are less likely to read entire journals; instead they conduct electronic literature searches on particular topics and find specific articles from a wide variety of journals. Hence, as long as the journal is listed in the main databases (e.g., Web of Science, Scopus, or Google Scholar) and papers are available online, they can be read and cited based on their own merits, unaffected by their journals' physical availability, reputation, or IF.

Hence, before the electronic age, the citation rate of any given paper and its journal's IF mutually reinforced each other. A journal's IF was (and still is) based on its individual papers' citation rates, and the citation rate of any individual paper was affected by its journal's circulation and availability, which depended on its IF. Now the former is still true, but if new practices of literature search and usage limit the effect of journal IF on paper citation rates, the correlation between paper citation rate and IF should be decreasing over time. Additionally, the proportion of highly cited papers coming from the highest IF journals should be diminishing over time, Here we examine whether this is indeed the case, and consider the implications to the continued use of the IF on the future of scientific publishing. Data for three groups of disciplines are presented: natural and medical sciences taken altogether, physics, and social sciences, from 1900 to 2011.

## Methods

We used Web of Science (WoS) data from Thomson Reuters from 1900 to 2011, covering all areas of natural sciences, medical sciences and social sciences. The dataset covers the *Century of Science* and *Century of Social Sciences datasets* from 1900-1944, and the Science Citation Index (Expanded), the Social Sciences Citation Index and the Arts and Humanities Citation Index from 1945-2011 period. The disciplinary classification of journals used for natural and medical sciences in general, and of physics and social sciences in particular, is an adaptation of the classification used by the U.S. National Science Foundation (NSF), which categorizes each journal in only one discipline and specialty.

The dataset included 25,569,603 natural and medical sciences papers 3,211,026 physics papers and 879,494 social sciences papers. The total number of cited references analysed was 819,369,970. Humanities papers were excluded from the analysis because of their long citation windows and high uncitedness rates (Larivière, Gingras, & Archambault, 2009), but citations in humanities journals were included. To be included in the analysis, papers had to be published in a journal for which an IF could be calculated. However, references made by excluded papers were considered as citations for other papers included in the analysis. Some journals on some years did not have an IF, either because their papers did not receive any citations during the 2-year citation window, or because 2 years must elapse before new journals receive their first yearly IFs.

Given that Thomson Reuters does not compile IFs for the entire period studied, and that the exact method by which it calculates IFs is not entirely clear, and hence irreproducible (Moed & Van Leeuwen, 1995; Rossner, Van Epps, & Hill, 2007), the IF of each journal covered in the database was recalculated. In the Thomsom Reuters IF, some types of publications are used to count citations (the nominator), but do not themselves count as "papers" (the denominator). Here, IF was calculated the same way as the Thomson Reuters IF, except that (1) this asymmetry between the numerator and the denominator was corrected; except as noted above in regards to papers in the humanities, if papers were counted, their citations were also counted, and (2) citations to individual papers were counted during the entire 2 years following their respective publication year. Hence, IF data was not available for the first 2 years, and full 2 year citation windows were not possible for the last 2 years, leaving a complete dataset of both IFs and citation rates from 1902 to 2009. A large proportion of papers remained completely uncited at the beginning of the period (Larivière et al., 2009); to reduce their weight in the IF-citations relationship, the analyses were also conducted excluding uncited papers, both in the calculation of citation rates and IFs.

Two indicators were used to measure the strength of the relationship between IF of journals and citations of papers. The first indicator was the coefficient of determination ($r^2$). Each paper was assigned the IF of the journal in which it was published and the citations it received during two years following its publication year, and the $r^2$ between the two series of variables was calculated for each year. The second indicator is the yearly percentage of the most highly cited papers published in the most highly cited journals, and the yearly percentage of the most highly cited papers not published in the most highly cited journals.

## Results

Figures 1, 2 and 3 present the $r^2$ between IF and paper citations from 1902-2009, for all disciplines of the natural sciences and medical sciences together (Fig. 1), physics (Fig. 2) and social sciences (Fig. 3). For descriptive purposes, in cases where there was a clear break, these data were split into 1902-1958, 1959-1990 and 1991-2009. For medical and natural sciences (Fig. 1), there was an increase of the correlation between IF and paper citation rates from 1902 until the end of the 1990s. The strength of the relationship between IF and citations did not increase steadily throughout the 20th century. Two dips occurred after the two World Wars, likely as a result of changes in the research system. More interestingly and in contrast to the general pattern throughout most of the $20^{th}$ Century, since scientific information began to be disseminated

electronically, around 1990, the relationship between the IF and citation rates has been weakening.

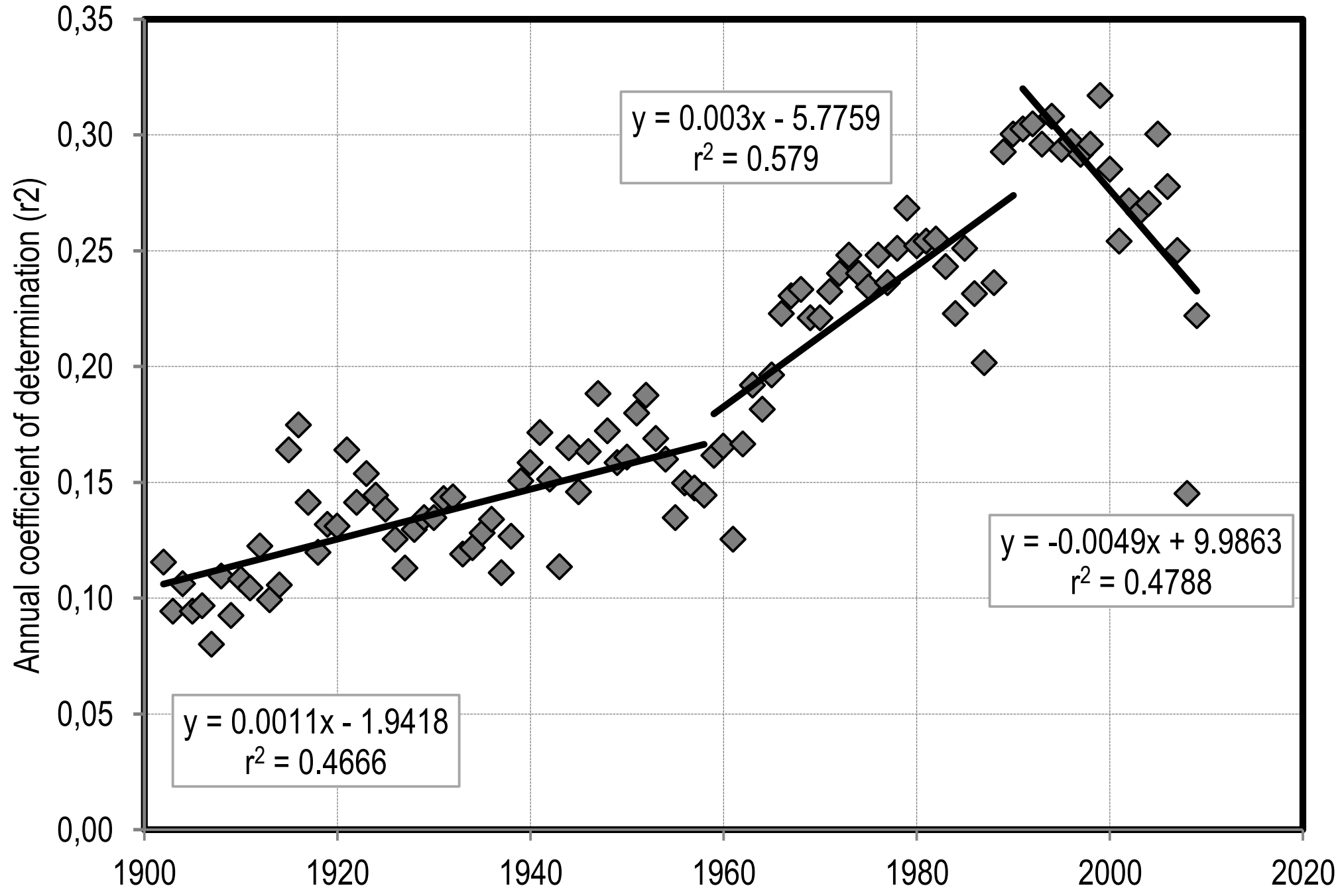

FIG. 1. Coefficient of determination ($r^2$) between the impact factor of journals and the 2-year citation rate of their papers from 1902 to 2009, for all natural and medical sciences journals.

The same analyses were carried out with 2 disciplines thought to be at opposite ends of the spectrum of how quickly they made the transition into the electronic domain: physics (Fig. 2) and social sciences (Fig. 3). Given the smaller sample size, the variation of the $r^2$ values between IF and papers' citation rates is larger, but in both cases there was a decrease during the last two decades. Although the decrease is not significantly different in the two disciplines, in physics it appears to start earlier, towards the end of the 1980s (Fig. 2).

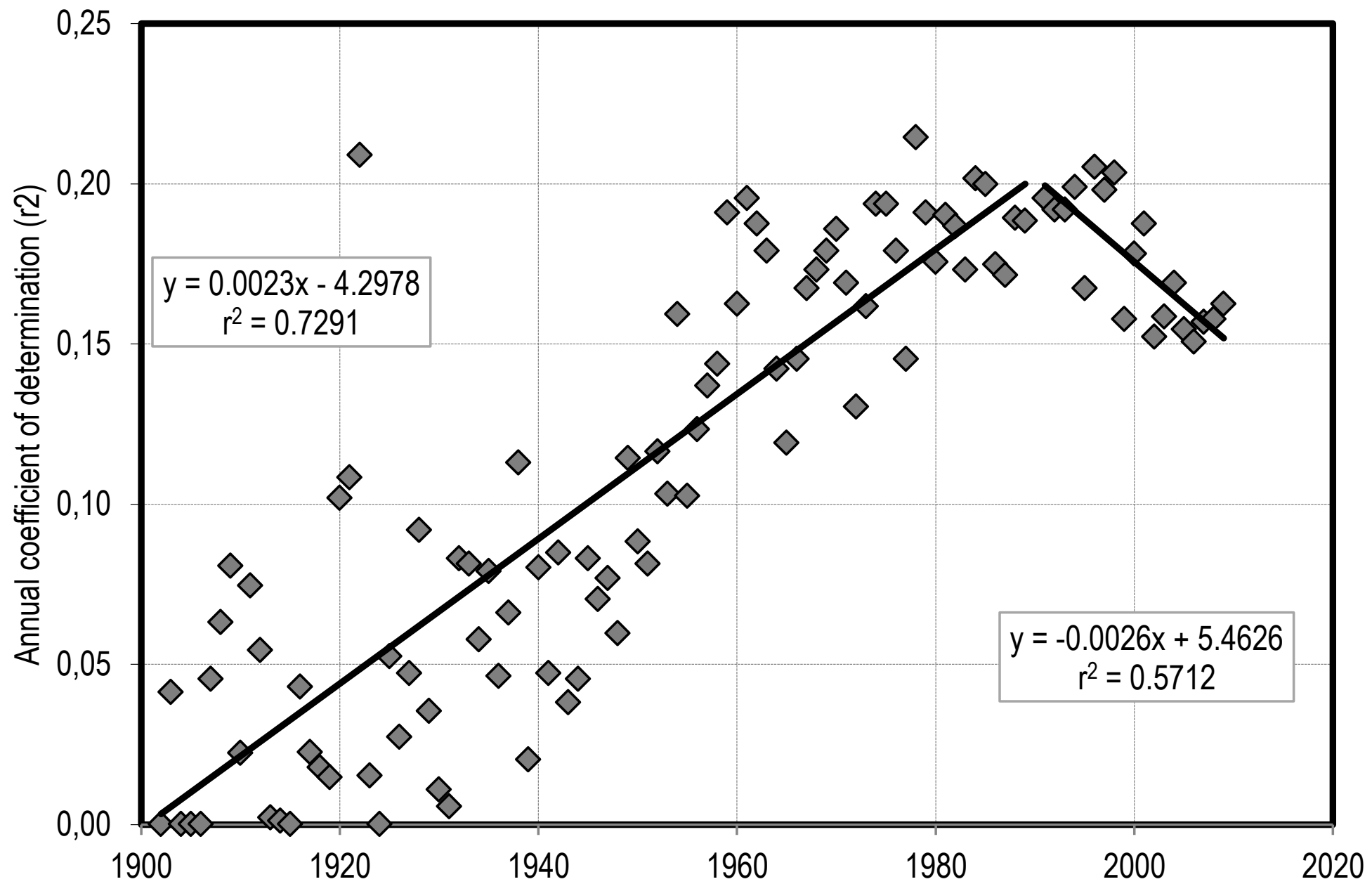

FIG. 2. Coefficient of determination ($r^2$) between the impact factor of physics journals and the 2-year citation rate of their papers from 1902 to 2009.

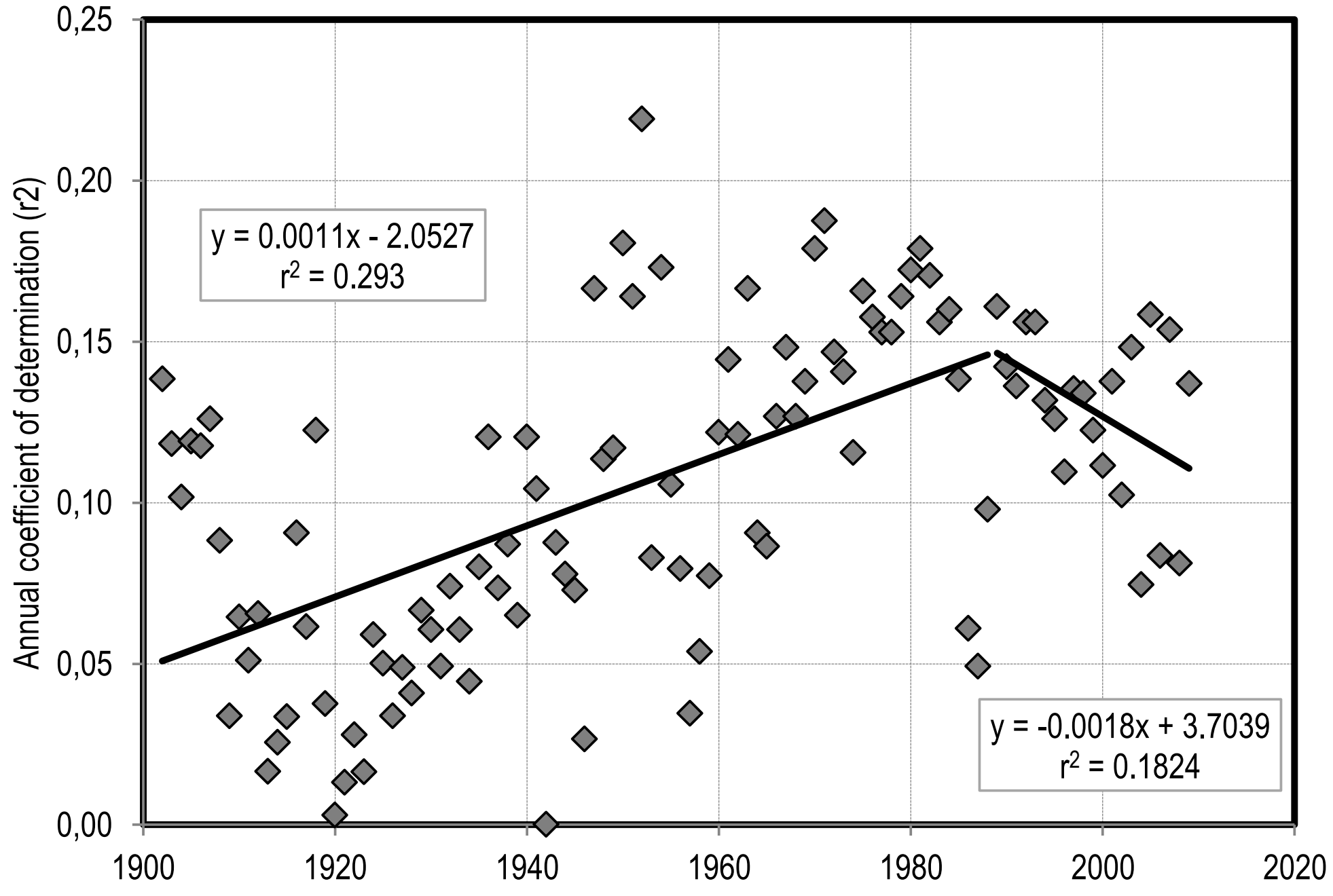

FIG. 3. Coefficient of determination ($r^2$) between the impact factor of social sciences journals and the 2-year citation rate of their papers from 1902 to 2009.

Although not shown, all results are similar and conclusions the same using Pearson's r and Spearman's rank correlation as indicators. All analyses were also carried out excluding uncited papers, both at the level of papers and in the calculation of the journal IFs. When uncited papers are excluded, a clearer trend with fewer fluctuations emerges, but the strength of the relationship between IF and citations remains within the same order of magnitude. So, removing uncited papers does not result in a stronger relationship between the IF and citations.

Figures 4 and 5 present an additional indicator of the relationship between IF and paper citations for all disciplines in the natural and medical sciences: the percentage of papers that are both in the top 10% (and 5%) most cited and published in the top 10% (and 5%) highest IF journals (Panels A of Figs. 4 and 5). By contrast, Panels B of these figures show the percentage of the top 10 (and 5%) most papers that are **not** published in the top 10% (and 5%) highest IF journals. Both figures show that the relationship between IF and citations has been weakening steadily since 1990, as a larger proportion of top (5 and 10%) most cited papers are published outside journals with top (5 and10%) IF.

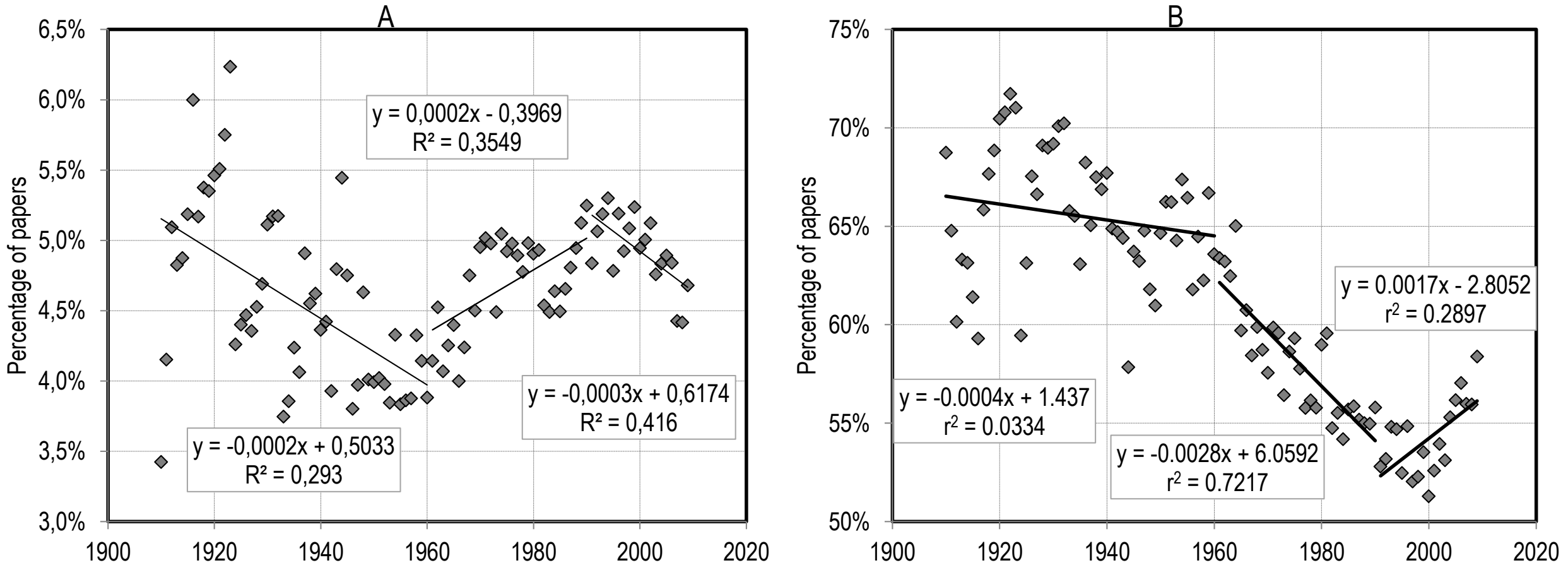

FIG. 4. A) Percentage of the top 10% most cited papers published in the top 10% most cited journals. B) Percentage of top 10% most cited papers that were **not** published in the top 10% most cited journals.

More specifically, the percentage of the 10% most cited papers published in the 10% most cited journals has been decreasing since 1990 (Fig. 4a), from about 5.25% to 4.50%. Accordingly, the percentage of the 10% most cited papers **not** published in the 10% journals with the highest IF has been increasing since 1990 (Fig. 4b); from 52% to about 56%. This pattern is even more clearly evident when the same comparisons are made for the top 5% of papers and the top 5% of journals (Fig. 5). In 1990 about 2.25% of the top 5% papers were published in top 5% journals; but by 2009 this figure had fallen to 1.90 % (Fig. 5a). Similarly, in 1990 about 55% of top 5% most cited papers were **not** published in the top 5% journals, but by 2009 the figure had increased to 62% (Fig. 5b). Hence, the most important literature is increasingly coming from a greater range of journals, not only the journals with the highest IF.

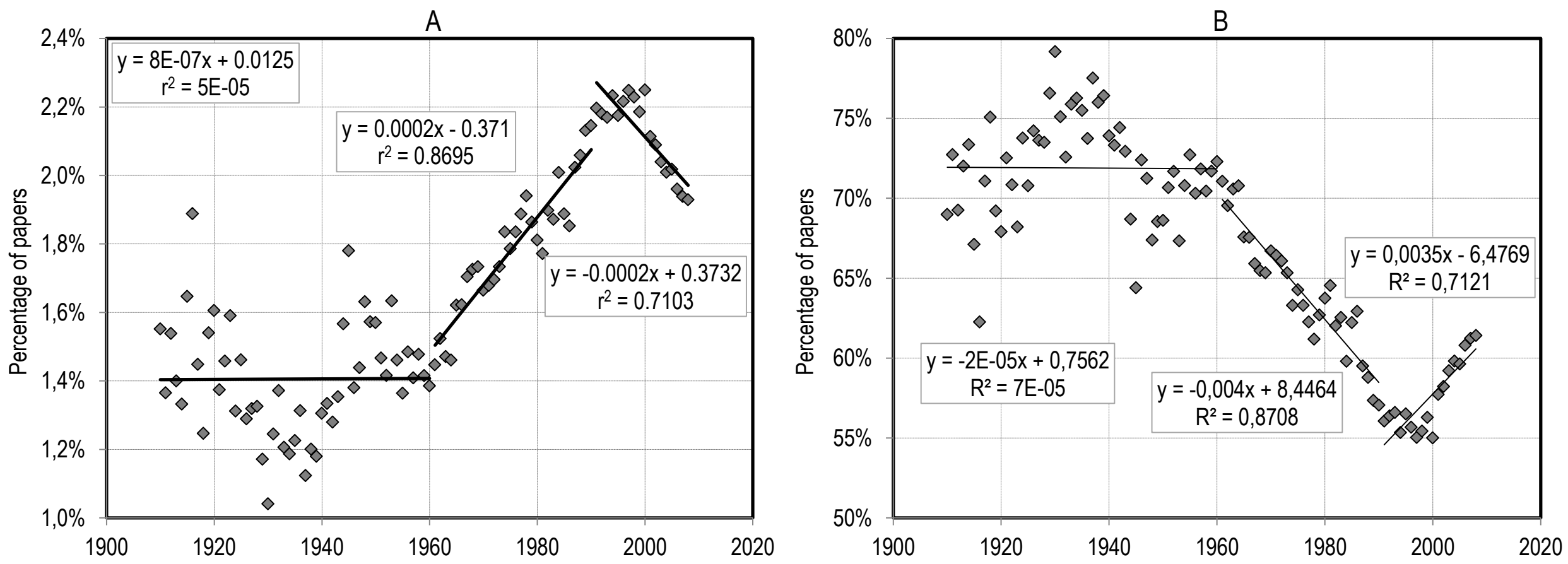

FIG. 5. A) Percentage of the top 5% most cited papers published in the top 5% most cited journals. B) Percentage of top 5% most cited papers that were **not** published in the top 5% most cited journals.

## Discussion and conclusions

Impact factors were initially developed to assist libraries in their purchasing decisions, and hence they have had a significant effect on journal circulation and availability. Here we show that throughout most of the 20th century, the link between the IF and papers' citations was getting stronger, but, as predicted, this link has been weakening steadily since the beginning of the electronic age. This change seems to have started slightly earlier in physics, a field that was quicker to adopt electronic dissemination of information. Also during this time the percentage of top papers coming from the top journals has been decreasing. Compounded with the fact that, in general, citations have become more widely spread among journals (Larivière et al., 2009), the electronic age and its new modes of disseminating and accessing scientific literature might bring forth the end of the IF as a useful measure of the quality of journals, papers, and researchers and have interesting implications for the future of scientific literature.

The IF has been repeatedly criticized as a suitable measure of journal quality (Aksnes, 2003; Morgan & Janca, 2000; Rossner et al., 2007; Rothenberg, 2008; Seglen, 1997; Vakil, 2005; Whitehouse, 2001). The strongest arguments against its validity and use are: (1) some types of publications within journals, such as letters and commentaries, are used to count citations (the nominator), but do not themselves count as "papers" (the denominator), and hence inflate the journal's IF, (2) the IF depends on the number of references, which differs among disciplines and journals, (3) the inclusion of journals in the database depends solely on Thomson Reuters, a private company, and not on the fields' practitioners, (4) the exact IF published by Thomson Reuters cannot be replicated using publicly available data, (5) the distribution of citations/paper is not normal, so at the very least the mode or median ought to be used instead of the mean, (6) the 2-year span for papers followed by one year for citations is completely arbitrary and favours high-turnover over long-lasting contributions, and (7) journal editors can manipulate and artificially inflate their IFs. Our analysis identifies one more problem: the relationship between paper quality and IF is weakening, so the IF is losing its significance as a measure of journal quality.

Second, IFs are used as a proxy for paper quality. Except for the most recently published papers that have not had a chance to be cited yet, there is no reason to use the IF as a proxy for a paper's quality. One can readily have access to any individual paper's citation rate and determine how the paper stands on its own, regardless of its journal's IF. As the relationship between paper citation rates and IF continues to weaken, and as more important papers increasingly appear in more diverse venues**,** it will become even less justifiable to automatically transfer a journal's reputation and symbolic capital on to even its most recently published papers. This should force a return to direct assessments of paper quality, by actually reading them.

Third, and even more troubling, is the 3-step approach of using the IF to infer journal quality, extend it to the papers therein, and then use it to evaluate researchers. Our data shows that the high IF journals are losing their stronghold as the sole repositories of high quality papers, so there is no legitimate basis for extending the IF of a journal to its papers, and much less to individual researchers. This is congruent with the finding that over the past decade in economics, the proportion of papers in the top journals produced by people from the top departments has been decreasing (Ellison, 2011). Moreover, given that researchers can be evaluated using a variety of other criteria and bibliometric indicators (e.g., Averch, 1989; Leydesdorff & Bornmann, 2011; Lozano, 2010; Lundberg, 2007; Põder, 2010), evaluating researchers by simply looking at the IFs of the journals in which they publish is both naive and uninformative.

For the past few centuries journals were a convenient way to organize papers by subject, but search engines now allow us to find individual papers on specific topics from across the entire spectrum of journals, so highly subject-specific journals might become obsolete or begin to amalgamate. Online, open-access journals, such as in the PLoS family of journals, and online databases, such as the ArXiv system and its cognates, will continue to gain prominence. Using these open-access repositories, experts could find publications in their respective fields and decide which ones are worth reading and citing, regardless of the journal. If this pattern continues, the IF will cease to be a meaningful measure of the quality of journals, papers and researchers.

## Acknowledgments

We thank Jean-Pierre Robitaille for comments on an earlier draft of this paper. G.A.L. thanks the University of Tartu for allowing him free access to their online collections.